\documentclass[twocolumn,showpacs,preprintnumbers,amsmath,amssymb]{revtex4}

\usepackage{graphicx}% Include figure files
\usepackage{dcolumn}% Align table columns on decimal point
\usepackage{bm}% bold math

\begin{document}

\title{Relativistic unitary coupled cluster theory and applications}

\author{Chiranjib Sur, Rajat K Chaudhuri, Bijaya K. Sahoo, B. P. Das}
\affiliation{Non-Accelerator Particle Physics Group, Indian Institute of Astrophysics,
Bangalore - 560 034, India}

\author{D. Mukherjee}

\affiliation{Department of Physical Chemistry, Indian Association for the Cultivation of Science, Kolkata -
700 032, India}

\date{December 14, 2004}

\begin{abstract}
We present the first formulation and application of relativistic unitary coupled
cluster theory to atomic properties. The remarkable features of this theory are
highlighted, and it is used to calculate the lifetimes of $5^{2}D_{3/2}$ and 
$6^{2}P_{3/2}$ states of $Ba^{+}$ and $Pb^{+}$ respectively. The results clearly
suggest that it is very well suited for accurate \emph{ab initio} calculations
of properties of heavy atomic systems.

\end{abstract}

\maketitle

There have been a number of attempts to modify  
coupled-cluster (CC) theory \cite{bishop},
despite its spectacular success in elucidating the properties of a wide range of
many-body systems \cite{bishop,lindgren-book,kaldor,bartlett-book,crawford}.
One interesting case in point is unitary coupled-cluster (UCC) theory which 
was first proposed by Kutzelnigg \cite{kutzelnigg1}. In this theory, the 
effective Hamiltonian is Hermitian  by construction and the energy which
is the expectation value of this operator in the reference state is unlike 
in CC theory,
an upper bound to the ground state energy \cite{crawford}. 
Another attractive feature of
this theory which we shall discuss later is that at a given level of
approximation it incorporates certain higher order excitations that are 
not present in CC theory. Furthermore, it is well suited for the calculation 
of properties where core relaxation effects are important \cite{geetha-thesis,
csur-jpb}. 
In spite of the aforementioned advantages, there have been 
relatively few studies
based on this method \cite{bartlett-cpl-89-1,ucc-hoffmann}. This
work is the first relativistic formulation of unitary coupled
cluster (UCC) theory and also the first application 
of this theory to atomic properties.
In this letter, we first present the formal aspects of relativistic UCC 
theory and then apply it to calculate the lifetimes of
the $5^{2}D_{3/2}$ and $6^{2}P_{3/2}$ states of $Ba^{+}$ and $Pb^{+}$ 
respectively; which depend strongly on both relativistic and 
correlation effects. The comparison of the results of these calculations with accurate experimental data would constitute an important test of this theory.

The exact wave function for a closed shell state in CC theory is obtained 
by the action of the
operator $\exp(T)$ on the reference state $\left|\Phi\right\rangle $.
However, in  UCC theory \cite{ucc-hoffmann}, it is written as
\begin{equation}
\left|\Psi\right\rangle =\exp(\sigma)\left|\Phi\right\rangle \label{eqn-1}
\end{equation}
where $\sigma=T-T^{\dagger}$; and $T$ and $T^{\dagger}$ are the excitation 
and deexcitation operators respectively.
$\sigma$ is clearly
anti-Hermitian, since $\sigma^{\dagger}=-\sigma.$
Using this unitary ansatz for the correlated wave function, the relativistic
UCC equation in the Dirac-Coulomb approximation can be written as 
\begin{equation}
\exp(\sigma^{\dagger})H\exp(\sigma)\left|\Phi\right\rangle =E\left|\Phi\right\rangle ,\label{ev-eqn}
\end{equation}
where $H$ is the Dirac-Coulomb Hamiltonian
\begin{equation}
H=\sum_{i}c\alpha_{i}\cdot p_{i}+(\beta-1)mc^{2}+V_{N}+\sum_{i<j}\frac{e^{2}}{r_{ij}}.\label{dirac-coul}
\end{equation}

Using the normal ordered Hamiltonian, Eq.(\ref{ev-eqn}) can
be rewritten as
\begin{equation}
\exp(\sigma^{\dagger})H_{N}\exp(\sigma)\left|\Phi\right\rangle =\Delta E\left|\Phi\right\rangle ,\label{ev-norm-eqn}\end{equation}
where the normal ordered Hamiltonian is defined as $H_{N}=H-\left\langle \Phi\right|H\left|\Phi\right\rangle $ and $\Delta E=E-\left\langle \Phi\right|H\left|\Phi\right\rangle $.
.

The choice
of the operator $\sigma$ makes the effective Hamiltonian $\overline{H}_{N}=\exp(-\sigma)H_{N}\exp(\sigma)$ 
Hermitian.

The effective Hamiltonian  is expressed by the Hausdorff expansion in CC theory
as
%%%%%%%%%%%%%%%%%%%%%%%%%%%%%%%%%%%%%%%%%%%%%%%%%%
\begin{eqnarray}
\overline{H}_{N}=&&\exp(-T)H_{N}\exp(T)\nonumber\\
&&
=H_{N}+\left[H_{N},T\right]
+\frac{1}{2!}\left[\left[H_{N},T\right],T\right]\nonumber\\
&&
+\frac{1}{3!}\left[\left[\left[H_{N},T\right],T\right],T\right]\nonumber\\
&&
+\frac{1}{4!}\left[\left[\left[\left[H_{N},T\right],T\right],T\right],T\right].
\label{h_n-ccsd}
\end{eqnarray}
%%%%%%%%%%%%%%%%%%%%%%%%%%%%%%%%%%%%%%%%%%%%%%%%%%
In UCC, the operator $T$ is replaced by $\sigma=T-T^{\dagger}$ in the above
equation and this results in $\overline{H}_{N}$ being expressed in terms of
a non-terminating series of commutators.
For practical  reasons, one  
truncates the series after some finite order. Truncation at the \emph{n}-th
order commutator leads to the nomenclature UCC(\emph{n}). 

Using UCC(\emph 3) approximation and without modifying the last term of Eq. (\ref{h_n-ccsd}), we write Eq. (\ref{ev-norm-eqn}) as

\begin{widetext}
\begin{equation}
\left[H_{N}+\overline{H_{N}T}+\frac{1}{2!}\left(\overline{\overline{H_{N}T}T}+2\overline{\overline{T^{\dagger}H_{N}}T}\right)\right.
\left.+\frac{1}{3!}\left(\overline{\overline{\overline{H_{N}T}T}T}+3\overline{T^{\dagger}\overline{T^{\dagger}\overline{H_{N}T}}}+3\overline{T^{\dagger}\overline{\overline{H_{N}T}T}}\right)
+\frac{1}{4!}\overline{\overline{\overline{\overline{H_{N}T}T}T}T}\right]\left|\Phi\right\rangle  
=\Delta E\left|\Phi\right\rangle 
\label{ucc3-eqn}
\end{equation}
\end{widetext}

\begin{figure*}
\begin{center}
\begin{tabular}{cc}
\includegraphics[%
 scale=0.7]{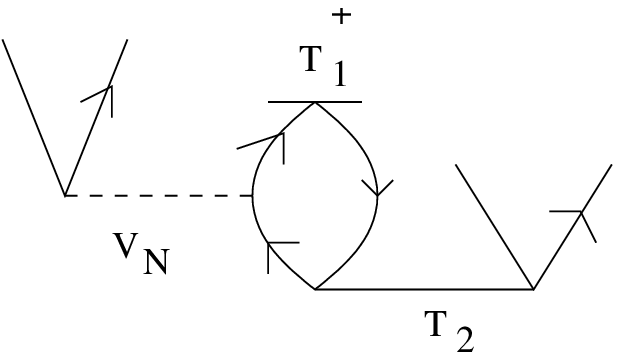}&
\includegraphics[%
 scale=0.7]{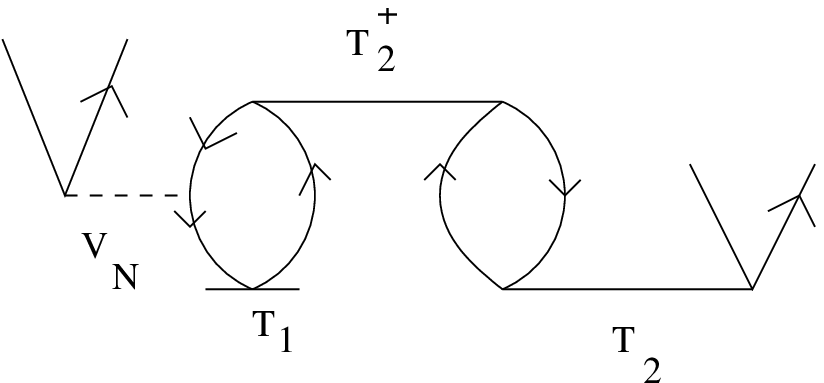}
\tabularnewline
(a)&
(b)
\tabularnewline
\end{tabular}
\end{center}
\caption{\label{fig-triples}Typical triples quadruples diagrams arising 
from UCC(\emph{3})}
\end{figure*}

By projecting single, double and higher order excited determinant states on
Eq.(\ref{ucc3-eqn}), we get the cluster amplitude equations.
The approximation which includes only single and double 
excitations/deexcitations in the UCC wavefunction is known as UCCSD.
After careful analysis, one finds that there are new terms arising from 
 $\overline{\overline{T^{\dagger}H_{N}}T}$,
$\overline{T^{\dagger}\overline{T^{\dagger}\overline{H_{N}T}}}$
and $\overline{T^{\dagger}\overline{\overline{H_{N}T}T}}$
in the UCCSD equations. 
The first term will give rise to some extra diagrams which correspond 
to double and triple excitations and the last two terms to triple
and quadrupole excitations. Some typical diagrams which represent 
triple and quadrupole excitations are given in figures
\ref{fig-triples}a and \ref{fig-triples}b respectively. 
%%%%% Figure 1 -- for triples%%%%%%%%%%%%%%%%%
%%%%%%% Figure 1 ends here %%%%%%%%%%%%%%%%%%%%%
The UCCSD approximation in addition to single and double excitations, also 
includes some triple and quadrupole excitations to all orders in the residual
Coulomb interaction in a more elegant and simpler manner compared to CC theory. 
It would have been computationally prohibitive to include triple and 
quadrupole excitations for heavy atoms in the framework of normal CC theory.
Indeed one of the principal advantages of UCC theory is its ability to subsume
higher levels of excitations than CC theory at the same level of approximation.

To calculate the ground state of the system we first compute
the cluster amplitude of the closed shell states($Ba^{++}$ and $Pb^{++}$)
 by using Eq.(\ref{ucc3-eqn})
and then use the open shell coupled cluster method (OSCC)\cite{csur-mg+}.
The exact wave function can then be written as
\begin{equation}
\left|\Psi_{k}^{N+1}\right\rangle =\exp(\sigma)\left\{ 1+S_{k}\right\} \left|\Phi_{k}^{N+1}\right\rangle ,\label{exact-n+1}
\end{equation}
where $\left|\Phi_{k}^{N+1}\right\rangle $ is the Dirac-Fock reference
state which we get after adding an electron to the $k$th virtual orbital and 
$S_{k}$ is the corresponding excitation operator.
We obtain a set of equations
\begin{equation}
\left\langle \Phi_{k}^{N+1}\right|\overline{H}_{N}\left(1+S_{k}\right)\left|\Phi_{k}^{N+1}\right\rangle =H_\mathrm{{eff}}\label{open-1}
\end{equation}
 and
\begin{eqnarray}
&&\left\langle \Phi_{k}^{^{\star}N+1}\right|\overline{H}_{N}\left(1+S_{k}\right)\left|\Phi_{k}^{N+1}\right\rangle \nonumber\\
&&=\left\langle \Phi_{k}^{^{\star}N+1}\right|S_{k}\left|\Phi_{k}^{N+1}\right\rangle H_\mathrm{{eff}}.
\label{open-2}
\end{eqnarray}
The Eq.(\ref{open-2}) is non-linear in $S_{k}$ because $H_\mathrm{{eff}}$
is itself a function of $S_{k}$ where 
$S_{k}=\sum_{ak}s_{a}^{k}a_{k}^{\dagger}a_{a}+
\sum_{abkr}s_{ab}^{kr}a_{k}^{\dagger}a_{r}^{\dagger}a_{b}a_{a}$;
 $s_{a}^{k}$ and $s_{ab}^{kr}$ are the single and double excitation cluster 
amplitudes for the valence electrons. The labels $a,b$ and $k,r$ refer to the
core and virtual orbitals respectively. Hence, these equations have to be 
solved self-consistently to determine the $S_{k}$ amplitudes. We have 
included the triple excitations in our calculations in an approximate way. 
The amplitudes corresponding to these excitations are of the form
\begin{equation}
S_{abk}^{pqr}=\frac{\widehat{VT_{2}}+\widehat{VS_{2}}}{\varepsilon_{a}+\varepsilon_{b}+\varepsilon_{k}-\varepsilon_{p}-\varepsilon_{q}-\varepsilon_{r}},\label{par-triples}
\end{equation}
where $S_{abk}^{pqr}$ are the amplitudes corresponding to the simultaneous
excitation of orbitals $a,b,k$ to $p,q,r$ respectively and $\widehat{VT}$
and $\widehat{VS}$ are the correlated composites involving $V$ and
$T$, and $V$ and $S$ respectively. $\varepsilon$'s are the orbital
energies. 
The above amplitudes are added appropriately in the singles and doubles open 
shell cluster amplitude equations and they are then solved 
self-consistently. We therefore obtain solutions of  $S_{1}$ and $S_{2}$  
amplitudes taking into consideration the partial effect of the triple 
excitations. We had referred to this approximation earlier as CCSD(T). 
%\cite{csur-mg+}.

As we have seen in Eq.(\ref{ucc3-eqn}), $\overline{H}_{N}$ is defined
as $\overline{H}_{N}=\exp(-\sigma)H_{N}\exp(\sigma)$. From Eq.(\ref{h_n-ccsd})
it is clear that the the expansion of $\overline{H}_{N}$ corresponding
to UCC theory can be expressed as

\begin{equation}
\overline{H}_{N}=(\overline{H}_{N})_\mathrm{{CCSD}}+(\overline{H}_{N})_\mathrm{{extra}}.\label{ucc-hnbar}\end{equation}
Following much the same argument as in the closed shell case, it can be shown
that the contributions from the extra part arising in Eq.(\ref{ucc-hnbar})
contains terms like $\overline{\overline{T^{\dagger}H_{N}}T}$,
$\overline{T^{\dagger}\overline{T^{\dagger}\overline{H_{N}T}}}$
and $\overline{T^{\dagger}\overline{\overline{H_{N}T}T}}$. These
terms will affect the open-shell amplitude determining equations.
In the UCC theory for the closed shell, we have seen that although
there are some extra terms arising from the expansion of $\overline{H}_{N}$,
the contributing diagrams corresponding to those terms are not always
new. Hence we have taken into account those diagrams which are not
present in the CCSD approximation. 
It can be shown that the extra terms
arising from the inclusion on the operator $\sigma=T-T^{\dagger}$
give rise to some new terms but the corresponding diagrams
have already been taken into account through the
inclusion of partial triple amplitudes. In analogy with CCSD(T), we refer to
our approach as UCCSD(T). 

The normalized transition matrix element ($i\longrightarrow f$) due
to an operator $\widehat{O}$ is given by

\begin{equation}
\widehat{O}_{fi}=\frac{\left\langle \Psi_{f}^{N+1}\right|\widehat{O}\left|\Psi_{i}^{N+1}\right\rangle }{\sqrt{\left\langle \Psi_{f}^{N+1}\right|\left.\Psi_{f}^{N+1}\right\rangle \left\langle \Psi_{i}^{N+1}\right|\left.\Psi_{i}^{N+1}\right\rangle }}\nonumber\\
\label{tr-mat}
\end{equation}

\begin{figure*}
\begin{center}\begin{tabular}{cc}
\includegraphics[%
  scale=0.7]{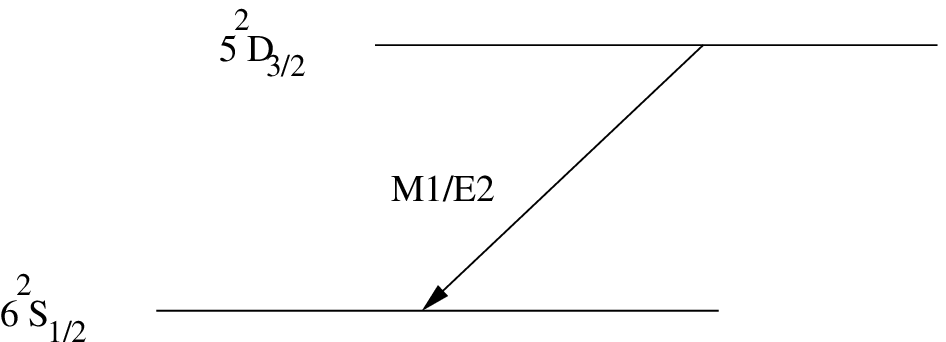}&
\includegraphics[%
  scale=0.7]{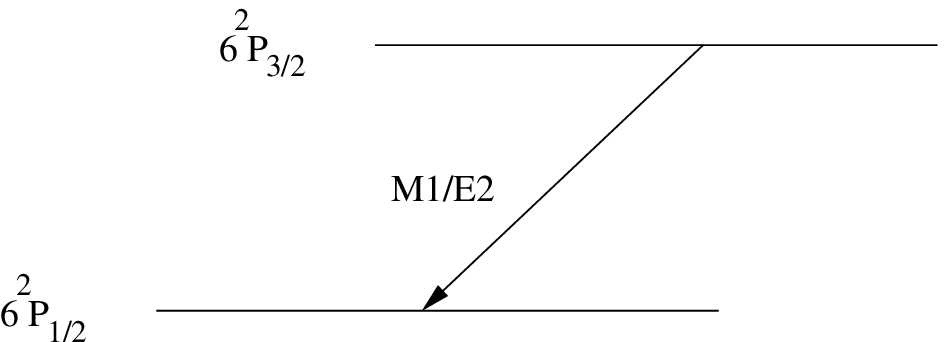}\tabularnewline
(a)&
(b)\tabularnewline
\end{tabular}\end{center}

\caption{\label{decay-scheme}Decay scheme of the low lying states of $Ba^{+}$and $Pb^{+}$} 
\end{figure*}

We consider next the calculations of the lifetimes of the $5^{2}D_{3/2}$ and 
$6^{2}P_{3/2}$ states of $Ba^{+}$ and $Pb^{+}$ respectively. Both of these 
states decay to the ground states via $M1$ and $E2$ transitions as shown in 
figure \ref{decay-scheme}.
The transition probabilities $A$ ($A_{fi}=A_{i\longrightarrow f}$
in $s^{-1}$) for $M1$ and $E2$ transitions are expressed as \cite{handbook}

\begin{equation}
A_{fi}^{M1}=\frac{2.6973\times10^{13}}{[J_{i}]\lambda^{3}}S_{fi}^{M1}\label{A-m1}\end{equation}
and\begin{equation}
A_{fi}^{E2}=\frac{1.1199\times10^{18}}{[J_{i}]\lambda^{5}}S_{fi}^{E2}\label{A-e2}\end{equation}
respectively, where $J_{i}$ is the degeneracy of the initial state and $\lambda$(in
\AA) is the wavelength corresponding to the transition $i\longrightarrow f$.
In Eqs.(\ref{A-m1}) and (\ref{A-e2}) the line-strength $S_{fi}$
is $\left|M1_{fi}\right|^{2}$ and $\left|E2_{fi}\right|^{2}$
respectively and $M1_{fi}$ and $E2_{fi}$ are the corresponding
one electron reduced matrix elements of magnetic dipole and electric
quadrupole transitions \cite{bijaya-pb+}.

The net probability for a given transition which allows two different
channels is given by \cite{handbook}

\begin{equation}
A=A_{M1}+A_{E2}\label{net-prob}\end{equation}
and the corresponding lifetime, which is the inverse of the transition
probability is expressed as 

\begin{equation}
\frac{1}{\tau}=\frac{1}{\tau_{M1}}+\frac{1}{\tau_{E2}}.\label{lifetime}\end{equation}

The results of the calculations of the lifetimes of the $5^{2}D_{3/2}$ state 
of $Ba^{+}$ and $6^{2}P_{3/2}$ state of $Pb^{+}$ are given in 
tables \ref{comp-table-ba+} and \ref{comp-table-pb+} respectively. These 
calculations are very challenging as they involve accurate determinations 
of the line strengths of the $M1$ and $E2$ transitions and more critically
the third and fifth powers of the excitation energies.

%
%%%%%%%%%%%%% Beginning of Table I%%%%%%%%%%%%%%%%%%%%%%%%%%%%%%%%%%%%%%%%
\begin{table}

\caption{\label{ba-basis}Description of the basis functions used in the UCC
calculation of $Ba^{+}$.}

\begin{center}\begin{tabular}{cccccccccc}
\hline 
&
$s_{1/2}$&
$p_{1/2}$&
$p_{3/2}$&
$d_{3/2}$&
$d_{5/2}$&
$f_{5/2}$&
$f_{7/2}$&
$g_{7/2}$&
$g_{9/2}$\tabularnewline
\hline
\hline 
Analytical &
5&
6&
6&
5&
5&
8&
8&
6&
6\tabularnewline
Numerical&
8&
7&
7&
5&
5&
2&
2&
0&
0\tabularnewline
Total&
13&
13&
13&
10&
10&
10&
10&
6&
6\tabularnewline
\hline
\hline 
&
&
&
&
&
&
&
&
&
\tabularnewline
\end{tabular}\end{center}
\end{table}
%%%%%%%%%%%%% End of Table I%%%%%%%%%%%%%%%%%%%%%%%%%%%%%%%%%%%%%%%%

%%%%%%%%%%%%% Beginning of Table II%%%%%%%%%%%%%%%%%%%%%%%%%%%%%%%%%%%%%%%%
\begin{table}

\caption{\label{comp-table-ba+}Excitation energies (in $cm^{-1}$) and lifetime 
(in $s$) of $5^{2}D_{3/2}$ state of $Ba^{+}$ corresponding to the transition 
$5^{2}D_{3/2}\longrightarrow 6^{2}S_{1/2}$}

\begin{center}\begin{tabular}{ccc}
\hline 
&
Excitation energy&
$\tau$\tabularnewline
\hline
\hline 
UCCSD(T)&
4789&
81.01\tabularnewline
CCSD(T) \cite{geetha-ba+}&
4809&
87.06\tabularnewline
Dzuba \emph{et al} \cite{dzuba}&
4411&
81.5\tabularnewline
Guet \emph{et al} \cite{johnson}&
4688&
83.7\tabularnewline
Experiment \cite{ba-exp}&
4874&
79$\pm$4.6\tabularnewline
\hline
\hline 
&
\tabularnewline
\end{tabular}\end{center}
\end{table}

%%%%%%%%%%%%% End of Table II%%%%%%%%%%%%%%%%%%%%%%%%%%%%%%%%%%%%%%%%
%

The lifetime of the  $5^{2}D_{3/2}$ state of $Ba^{+}$
is calculated using hybrid orbitals consisting of single particle 
orbitals which are partly numerical
and partly analytical. Such an approach is described in detail
by Majumder \emph{et al} \cite{hybrid}. The analytical orbitals are
Gaussian type orbitals (GTOs).
The details of this basis are given in table \ref{ba-basis}. Excitations from
all the core orbitals are included for this ion as well as $Pb^{+}$.
Since the calculation of transition probabilities/lifetimes are extremely
sensitive to the excitation energies. 
Gopakumar \emph{et al}
\cite{geetha-ba+} have shown that the hybrid basis can serve
this purpose very well. Our UCCSD(T) calculations yield
ionization potentials (IPs) to an accuracy of about 0.1\%,
and 0.07\% for the $6^{2}S_{1/2}$ and $5^{2}D_{3/2}$ 
states of $Ba^{+}$ respectively. The corresponding IP (in $cm^{-1}$)
values are 80544 (80687) and 75755 (75813). The numbers
in the parentheses are the corresponding experimental values 
\cite{moore-tables}. 

%%%%%%%%%%%%% Beginning of Table III%%%%%%%%%%%%%%%%%%%%%%%%%%%%%%%
\begin{table}

\caption{\label{comp-table-pb+}Excitation energies (in $cm^{-1}$) and lifetime (in $s$) of the $6^{2}P_{3/2}$ state
of $Pb^{+}$ }

\begin{center}\begin{tabular}{ccc}
\hline 
&
Excitation energy &
$\tau$
\tabularnewline
\hline
\hline
\tabularnewline
Dirac-Fock&
13612 &
0.0415\tabularnewline
CCSD(T)\cite{bijaya-pb+}&
13710 &
0.0425\tabularnewline
UCCSD(T)&
13719 &
0.0413\tabularnewline
Exp\cite{pb+-exp}&
14085 &
.0412(7)\tabularnewline
\hline
\hline 
&
&
\tabularnewline
\end{tabular}\end{center}
\end{table}
%%%%%%%%%%%%% End of Table III%%%%%%%%%%%%%%%%%%%%%%%%%%%%%%%%%%%%%%%

The values of the calculated and experimental lifetimes are given
in table \ref{comp-table-ba+}. Dzuba \emph{et al}\cite{dzuba}, 
Guet \emph{et al} \cite{johnson} and Gopakumar \emph{et al} 
\cite{geetha-ba+} \emph{}have not considered the $M1$ channel in their 
calculations. 
Gopakumar \emph{et al} have used CCSD(T), but
Guet \emph{et al} and Dzuba \emph {et al} have used different variants of 
many-body perturbation theory (MBPT) with certain semi-empirical features to
calculate the transition amplitudes. In addition, the latter two have used
experimental values of the excitation energies to calculate the transition
probabilities/lifetimes. Our calculation based on UCCSD(T) is purely
\emph{ab initio} and takes into account the contribution of the $M1$ transition.
The transition amplitude in this approach includes some terms in addition to
those that appear in CCSD(T); and they arise due to the presence of the
deexcitation operator in the UCC wavefunction.
The result of the lifetime calculation considering only the $E2$ channel is
85.567. Inclusion of the $M1$ transition improves the total value (81.01
$s$) which is within the experimental error bar (79$\pm$4.6 $s$).

The UCCSD(T) calculations  of lifetime for the $6^{2}P_{3/2}$ state of $Pb^{+}$ are carried out using the same basis functions used by Sahoo \emph{et al} 
\cite{bijaya-pb+}.  The leading correlation contribution to the $M1$ and
$E2$ channel come from the core polarization and pair correlation effects.
It is evident from Table \ref{comp-table-pb+} that the accuracy of the 
calculation of  excitation energies using UCCSD(T) is even better than 
that obtained from CCSD(T).
This result and the improvements in the $M1$ and $E2$ transition 
amplitudes due to the UCC formulation coupled together give a value of the 
lifetime of the $6^{2}P_{3/2}$ state of $Pb^{+}$ that is clearly in better
agreement with the corresponding CCSD(T) calculation. This value is within 
the limits of the experimental error \cite{pb+-exp}, but this is not the 
case for CCSD(T).

In summary, we have applied the relativistic UCC theory for the first time to 
atomic properties. The results of our UCCSD(T) calculations of the lifetimes 
of the  $5^{2}D_{3/2}$ and $6^{2}P_{3/2}$ states of $Ba^{+}$ and $Pb^{+}$ 
respectively are in very good agreement with experiment and superior to those 
of all previous calculations. It indeed appears that this theory is capable of
yielding high precision results for a wide range of properties of heavy atomic 
systems including violation of parity and time-reversal symmetries. In 
addition, it would be worthwhile to explore its feasibility for the accurate 
determination of the properties of other many-body systems, particularly 
when it becomes necessary to go beyond the usual coupled cluster theory.

This work was supported by the BRNS for project no. 2002/37/12/BRNS. The 
computations were done on our group's Xeon PC cluster and Param Padma, the 
Teraflop Supercomputer in CDAC, Bangalore. We thank Professor 
G$\mathrm {\ddot u}$enther Werth 
for useful discussions.

\end{document}